\documentclass{article}
\usepackage{mathptmx} 
\usepackage{amsmath}
\usepackage{amsfonts}
\usepackage{amssymb}
\usepackage{datetime}
\usepackage{geometry}
\newtheorem{theorem}{Theorem}

\newcommand{\proof}{Proof:\ }
\newcommand{\qed}{$\Box$}
\numberwithin{equation}{section}
\usepackage[absolute]{textpos}
\usepackage{setspace}
\usepackage{hyperref}
\usepackage{graphicx}
\usepackage{array}
\usepackage[labelformat=empty]{caption}	
\usepackage{float}

\title{Four Algorithms on the Swapped Dragonfly}
\date{}
\begin{document}

	\maketitle
	\begin{centering}
		Richard Draper \footnote{Center for Computing Sciences, Institute for Defense Analyses, Bowie, MD}
		
		rndrape@super.org,rndrape@gmail.com

	\end{centering}

	\thispagestyle{empty}
	\vspace{3em}

\begin{abstract}
	A Swapped Dragonfly with $M$ routers per group and $K$ global ports per router is denoted $D3(K,M)$. A Swapped Dragonfly with $K$ and $M$ restricted is studied in this paper. There are four cases. In each case the resulting Swapped Dragonfly has a special utility:
	
	\begin{itemize}
	\item If $K=L^2$, an $LM \times LM$ matrix product can be performed in $LM$ rounds. An $n \times n$ matrix product may be performed in $n^2/LM$ rounds. Each round takes four router hops.
	\item If $K=ks$ and $M=ms$ an all-to-all exchange may be performed in $KM^2/s$ rounds. An all-to-all exchange of size $n\geq KM^2$ may be performed in $n^2/KM^2s$ rounds.
	\item If $K=2^k$ and $M=2^m$, a dilation three emulation of the $(k+2m)$ dimension Boolean hypercube exists. An ascend-descend algorithm may be performed at twice the cost on the hypercube.
	\item $D3(K,M)$ contains $M$ adjacent depth four edge-disjoint spanning trees. Equipped with a synchronized source-vector packet header, it can perform $n$ broadcast in $3n/M$ rounds.
\end{itemize}	
The rounds in these algorithms are free of link conflicts. These results are compared with algorithms on a hypercube. Comparisons with the fully populated Dragonfly are also made.These results are more applicable than the special cases because $D3(K,M)$ contains emulations of every Swapped Dragonfly with $J\le K$ and/or $L \le M$.

The Swapped Dragonfly is a new approach to the Dragonfly concept. It behaves much like a three-dimensional mesh. The underlying graph is defined before the network is defined.  The underlying graph of a Dragonfly is not defined until ports are identified on routers and links are defined by connections with ports. 
\end{abstract}

\section{Introduction}
This paper presents algorithms which can be performed on a Swapped Dragonfly interconnection network more efficiently than they can be performed on a fully-populated Dragonfly interconnection network. A knowledge of a Swapped Dragonfly \cite {Dra:swp} and a fully populated Dragonfly \cite{kim:dragonfly} is assumed. (The term "interconnection network" is elided hereafter.) A fully-populated Dragonfly will be called a Maximal Dragonfly. The information about a Maximal Dragonfly pertinent to this discussion is not generally available so explanations will be given when necessary.
 
The Swapped Dragonfly \cite{Dra:swp} is an interconnection network having $KM^2$ routers. Routers have coordinates $(c \bmod K, d \bmod M, p \bmod M)$. Connectivity is given by:
$$\begin{array}{clclclcl}
&&&&(c,d,p) &\stackrel{l}\longleftrightarrow &(c,d,p') &\textrm{ and}\\

&&&&(c,d,p) &\stackrel{g}\longleftrightarrow &(c',p,d)
\end{array}.$$
Note the swap of $d$ and $p$.
These are referred to as local and global connections, respectively. Conceptually the network is made up of $K$ cabinets containing $M$ drawers. Each drawer\footnote{A Dragonfly group is synonymous with a drawer.} has $M$ routers. There are $K$ global ports and $M-1$ local ports on each router. Connections are bidirectional. The local connections connect the routers in a drawer in a complete graph. The Swapped Dragonfly is denoted $D3(K,M)$.

The routers of a Maximal Dragonfly and a Swapped Dragonfly are identical. A Maximal Dragonfly has $KM^2+M$ routers. Routers have coordinates $(g,p), 0 \le g<KM+1$ and $1\le p < M$. The network is constructed from groups (a.k.a drawers) consisting of $M$  routers connected in a complete graph. The $K$ global ports of the routers are used to connect groups to one-another such that the groups are connected in a complete graph of order $KM+1$. The underlying graph of the network is a graph called a replacement graph \cite{Hoo:expan} of a complete graph of order $M$ in a complete graph of order $KM+1$                                                        . There are an enormous number of ways to construct a replacement graph \cite{Dra:fine}, but only one way has been used in the design of an actual machine \cite{alv:CRAY},\cite{Ari:percs}. This is the one to which the term Maximal Dragonfly in this paper refers. It is denoted $MDF(K,M)$.

$D3(K,M)$ is treated as a packet switching network using source-vector routing. At $(c,d,p)$ a source-vector $(\gamma,\pi,\delta)$ produces the path 

$$\begin{array}{clclclclclcl}
(c,d,p) &\stackrel{\delta}\longrightarrow &(c,d,p+\delta) &\stackrel{\gamma} \longrightarrow &(c+\gamma,p+\delta,d) &\stackrel{\pi}\longrightarrow & (c+\gamma,p+\delta,d+\pi)\\	
\end{array}$$

The following four properties of $D3(K,M)$ were established in \cite{Dra:swp} and will be referred to in this paper. 
\begin{enumerate}
	\item Simultaneously, every router can send a packet with header $(\gamma,\pi,\delta)$ without link conflicts. The result is a permutation of the routers.
	\item The routers of $D3(K,M)$ with $c$ in a set of size $J<K$ and $p$ in a set of size $L<M$ are connected in a closed sub-network isomorphic\footnote{Isomorphism is used here to mean dilation one emulation.} to $D3(J,L)$.
	\item If $\gamma \ne \gamma', \delta \ne \delta'$,and $\pi \ne \pi'$ two routers can simultaneously send packets with headers $(\gamma,\pi,\delta)$ and $(\gamma', \pi', \delta')$ without link conflicts.\label{Pro:3}
	\item $D3(K,M)$ scales linearly in $K$ and quadratically in $M$.
\end{enumerate}

This paper examines four special cases of the Swapped Dragonfly; $D3(K^2,M), D3(ks,ms)$, $D3(2^k,2^m)$ and $D3(K,M)$ with synchronizing header $[b;\gamma,\pi,\delta]$. Each is useful for a particular algorithm. On $D3(K^2,M)$ vector-matrix multiply takes one round of four hops with two off-and-ons to calculate products and accumulate sums. The algorithm can be in place or out of place. On $D3(ks,ms)$ an all-to-all exchange takes only $KM^2/s$ rounds. $D3(2^k,2^m)$ emulates a $k+2m$ Boolean-hypercube with maximal dilation three and average dilation two. $D3(K,M)$ with synchronizing header can perform $M$-broadcasts in one round of five router hops. $D3(2^k,2^m)$ with synchronizing header can emulate $(k+2m)$ Boolean hypercube with uniform dilation four.

Property 2 of $(D3(K,M)$ makes it possible to use these special Swapped Dragonflies "inside" a general $D3(K,M)$ for a range of values of $K$ and $M$ at a cost of some of the algorithm's performance.

\section{$D3(K^2,M)$}

$D3(K^2,M)$ may be viewed as a $K\times K$ array of $M\times M$ blocks. The index set for such an array is $\{(s,t,u,v), 0 \le s,\  t<K \textrm{ and } 0\le u,\ v < M\}$. The index set $(s,t,u,v)$ is assigned to $(c,d,p)=(s+tK,u,v)$. The arithmetic is done $\bmod \ K^2$ and canonical values are chosen as representatives of equivalence classes.  A $KM\times KM$ matrix  stored in this way can be transposed in a single global hop $\gamma=(s+th - (t+sK))$. 

A row vector $(s,u)$ refers to $(s,*,u,*)$ where $*$ denotes all possible values of the coordinate. The term vector will refer to the index set or to a $KM$-vector stored at the index set. A column vector $(t,v)$ refers to $(*,t,*,v)$. Note in $D3(K^2,M)$ a row vector $(s,u)$ is stored at (nodes attached to) $(s+*K,u,*)$ and a column vector $(t,v)$ is stored at $(*+tK,*,v)$. Note that a vector is stored on $KM$ routers.

\begin{theorem}
	On $D3(K^2,M)$ a $KM \times KM$ matrix product takes $KM$ rounds. Each round takes $4$ network hops and two off and ons.
\end{theorem}
\proof
To form the product, $VA$, of a row vector $V$ at $(s,u)$ and an $RM \times RM$ matrix $A$ on $D3(K^2,M)$ two phases are required. The first is to bring the row vector into juxtaposition with the columns of $A$ so that $V_{t,v}A_{t,t',v,v'}$ can be computed. The second phase is to accumulate these values at the row vector $(s,u)$. The result is an in-place algorithm.

The first phase is done by broadcasting element $(t,v)$ at row $(s,u)$ to all locations in row (t,v,) of $A$:

\begin{equation} (s+tk,u,v)\stackrel{G}\longrightarrow(t+*K,v,u)\stackrel{L}\longrightarrow(t+*K,v,*) \ \forall (t,v). \label{eq:one}\end{equation}
The first broadcast is over all global ports  of $(s+tk,u,v)$. The second broadcast is over all local ports of $(t+sK,v,u) \forall s$.  If $(t,v) \ne (t,v_1)$, there are no link conflicts in \ref{eq:one} because the center routers are all distinct. If $t \ne t_1$ the center routers are in different cabinets. If $v \ne v_1$ the center routers are in different drawers. Therefore, $KM$ broadcasts can occur simultaneously. The first phase is completed in two network hops. That is, row $(s,u)$ can be brought into juxtaposition with every column of $A$ in two network hops. The juxtaposed pairs hop off, are multiplied, and the product hops on. A path in \ref{eq:one} has the form:
\begin{equation}(s+tK,u,v)\stackrel{g}\longrightarrow (t + t' K,v ,u) \stackrel{l}\longrightarrow( t+t'K,v ,v'). \label{eq:two} 
\end{equation}
The accumulation phase of the algorithm requires $(t,t',v,v')$ to map to $(s,t,u,v)$ for all $(t',v')$. The path \ref{eq:three} reverses path \ref{eq:two}, 
\begin{equation}
(t+t'K,v,v')\stackrel{l}\longrightarrow(t+t'K,v,u)\stackrel{g}\longrightarrow(s+tK,u,v) \ \forall (t',v'). \label{eq:three} 
\end{equation}
If $(t_1,v_1) \ne (t,v)$ the path 
$$(t_1+t'K,v_1,v'_1)\stackrel{l}\longrightarrow(t_1+t'K,v_1,u)\stackrel{g}\longrightarrow(s+t_1K,u,v_1)$$ 
does not conflict with \ref{eq:three} because the center routers are distinct. Therefore, the path can be followed simultaneously for all $(t,v)$. $M$ values arrive at $(t+t'K,v,u)$ after the local step. They hop off, are accumulated and the partial sum hops on. $K$ partial sums arrive at $(s+tK,u,v)$ after the global step. They hop off and are accumulated. The value is $\sum_{t,v}V_{t,v}A_{t,t',v,v'}$ which is element $(t',v')$ of the vector-matrix product. A matrix multiply takes $KM$ rounds.

The vector-matrix multiply takes one  round consisting of four network hops and two off-and-ons to perform arithmetic. The time is $4t_w + 2t_s$, where $t_w$ denotes router latency and $t_s$ is time for the on-and-off. It is presented as an in-place algorithm. By modifying $s$ and $u$ in the last two hops, it can be converted to an out of place algorithm.
\qed

The expected situation for this algorithm is an $n$-vector $V$ and a $n\times n$ matrix $A$ with $n\geq KM$. Let $X=n/KM$. There is an $X$ subvector $V_{s,t,u,v}$ and an $X\times X$ submatrix $A_{t,t',v,v'}$ for all $(t,t',v,v'); (s,u)$ is fixed. The broadcast path \ref{eq:one} is used $X$ times to bring all $V_{t,v,i}$ into juxtaposition with the columns $A_{t,t',v,v',i,i'}$. The $X$ vector $V_{t,v,i}$ hops off the network so that the vector-matrix product $\sum_iV_{t,v,i}A_{t,t',v,v',i,i'}$ can be computed. (This was a scalar product when $n=KM$.) The result is $(VA)_{t,t',v,v',i'}$. For each $(t,t',v,v')$ it is an $X$ vector. These are accumulated using path \ref{eq:three} $X$ times for each $(t,t',v,v')$. The result is an $X$ vector at $(s,t,u,v)$. The $n$ vector $(s,*,u,*)$ of these $X$ vectors is $VA$. The vector multiply takes $n/KM$ rounds. The matrix multiply takes $n^2/KM$ rounds because the vector multiply must be used $n$ times. There is a cost for the $X\times X$ product which is independent of the network cost.

\begin{theorem}
On $D3(K^2,M)$ an $n \times n$ matrix product with $n >> KM$ takes $n^2/KM$ rounds. Each round takes $(4t_w+2t_s)$ time where $t_w$ is router latency and $t_s$ is time for off-and-on.
\end{theorem}

It is possible to transfer this algorithm to a $KM \times KM$ matrix on $D3(K,M)$. It requires storing $(s,t,u,v)$ at $(s,u,v)\  \forall t$. The memory requirements are increased by a factor of $K$ at every router. A row vector $(s,u)$ is stored as $K$-tuples at $(s,u,*)$ and a column vector $(t,v)$ is stored at $(*,*,v)$ but only one entry of the vector is at each $(c,d,v)$. A vector-matrix multiply takes $K$ rounds and a matrix multiply takes $K^2M$ rounds. 

If $(L+1)^2>K>L^2$, it is faster to do a vector-matrix multiply on $D3(L^2,M)$ than on $D3(K,M)$ because a vector of length $KM$ takes $K/L$ rounds on $D3(L^2,M)$ and $K$ rounds on $D3(K,M)$. This procedure is made possible by property $2$ of the Swapped Dragonfly.

The following table presents the network cost of matrix multiplication algorithms. The notation is $n\times n$ matrices with $P$ processors. The Cannon algorithm\cite{Can:alg} was originally done on a $\sqrt{P}\times\sqrt{P}$ mesh. The other algorithms are on a Boolean hypercube with $P$ nodes. DNS refers to Dekel, Nassimi, and Sahni \cite{Dekel:par}; HJE refers to Ho, Johnsson and Edleman \cite{Ho:hyper}; and GS refers to Gupta and Sudayappan\cite{Gup:hyper}. A version of the DNS algorithm appears in both HJE and GS. These algorithms divide the hypercube into $\sqrt{P} \times \sqrt{P}$ mesh. The $\log{P}$ term in two of the algorithms comes about by using an all-to-all algorithm due to Ho, Johnsson and Edelman\cite{Ho:hyper} which will be discussed in the last section. On $D3(L,M), P=LM^2$. The table shows that the result on $D3(K^2,M)$ is in keeping with results on other topologies. Only network costs appear and $t_w$ represents network latency.

$$
\begin{array}{llll}
\vspace{.05in}
&D3(K^2,M) &&4t_wn^2/\sqrt{P}\\
\vspace{.05in}
&\textrm{Cannon} &&2t_wn^2/\sqrt{P}\\
\vspace{.05in}
&\textrm{HJE} &&2t_wn^2/\sqrt{P}\ \log{P}\\
\vspace{.05in}
&\textrm{DNS} &&2t_wn^2/\sqrt{P}\\
\vspace{.05in}
&\textrm{GS} &&3t_wn^2/P^{2/3}\log{P}\\
\vspace{.05in}
&\textrm{DNS} &&4t_wn^2/P^{2/3}\\
\end{array}
$$

\section{$D3(ks,ms)$}
	
	\begin{theorem}
	On the Swapped Dragonfly, $D3(ks,ms)$, an all-to-all exchange among $n\geq KM^2$ nodes can be performed in $n^2/KM^2s$ rounds.
	\end{theorem}  
		
	\proof
	In $\boldsymbol{Z} \bmod M, \  s$ generates a subgroup $G=\{0,s,\cdots,(m-1)s \}$. $G$ has $s$ cosets. Denote them $[0],\cdots,[s-1]$. An analogous statement applies to $K$. An example is instructive. If $m=5 \textrm{ and } s=3 \textrm{ then }G=\{0,3,6,9,12\} \textrm{ and a coset } [t] \textrm{ is } t+G$;
	$$
	\begin{array}{llcc}
	$$
	&[0] &= &\{0,3,6,\hspace{1mm}9,12\}\\
	&[1] &=&\{1,4,7,10,13\}\\
	&[2] &=&\{2,5,8,11,14\}
\end{array}
$$
Notice that the cosets partition $\boldsymbol{Z} \bmod M$ into three disjoint sets and that the columns partition it into five disjoint sets. The second partition is called a \emph{dual} partition.

Consider the following array which is called a \emph{disagreeable array} (DA).

$$\begin{array}{ccccccc}
i &&0 &1 &\cdots &s-1\\
\gamma && [0] & [1] & \cdots & [s-1]\\
\pi && [0] & [1] & \cdots & [s-1]\\
\delta && [0] & [1] & \cdots & [s-1]
\end{array}$$
Each column contains $km^2$ vectors $(\gamma,\pi,\delta); k \textrm{ values of } \gamma, m \textrm{ values of }\pi,\textrm{and }m \textrm{ values of }\delta$. If $i \ne j$ and {$(\gamma,\pi,\delta)$} is a vector from column $i$ and $(\gamma ', \pi ' , \delta ')$ is a vector from column $j$ then $\gamma \ne \gamma ', \pi \ne \pi ', \textrm{ and } \delta \ne \delta '$. Therefore, every router can simultaneously send packets on the vector paths $(\gamma,\pi,\delta)$ and $(\gamma ',\pi ',\delta')$ without link conflicts by property \ref{Pro:3}. If one vector is taken from each column of the array, every router can simultaneously send $s$ packets without link conflict. This is denoted $(lgl)^s$ and is a round of the algorithm being developed here. It takes $3$ hops. There are $km^2$ vectors in each column of the DA. Therefore, there are $km^2$ rounds $(lgl)^s$ delivering $km^2s$ packets. Note that each vector is used only once in this process.

Cyclically shifting row $\pi$ one place to the left produces a new DA. No vector in the new array appeared in the previous array. Suppose $(\gamma,\pi,\delta)$ is in the original array and $(\gamma',\pi',\delta')$ is in this array. If $\pi = \pi', (\gamma,\delta) \ne (\gamma', \delta')$ because $\pi$ and $\pi'$ are in the same coset so $(\gamma,\delta)$ and $(\gamma',\delta')$ are in different cosets. If $(\gamma,\delta) = (\gamma', \delta')$ then $\pi$ and $\pi'$ are in different cosets. Therefore, this new array produces $km^2$ rounds $(lgl)^s$ delivering $km^2s$ packets to a different set of destinations. A series of $s^2$ left shifts of the bottom two rows produces $s^2$ distinct DAs. Each yields $km^2s$ vectors in $km^2$ rounds $(lgl)^s$. The entire set is $km^2s^3 = KM^2$ in $KM^2/s$ rounds. If $n=XKM^2, K=ks$ and $M=ms$, an all-to-all exchange between $n$ nodes take $X^2KM^2/s = n^2/KM^2s$ rounds.
\qed 

To convert this argument into an algorithm it is necessary to specify the order in which elements are chosen from the cosets in a $DA$. A single column of the first DA has the form:
$$
\begin{array}{clcccccc}
&\gamma &[i] &i &i+s, &\cdots, &i+(k-1)s \bmod K\\
&\pi &[j] &j &j+s, &\cdots, &j+(m-1)s \bmod M\\
&\delta &[k] &k &k+s, &\cdots, &k+(m-1)s \bmod M
\end{array}
$$
The left shifts are determined by $\phi =\mu+\nu s, 0 \le \phi  < s^2$. Row $\pi$ of the array is shifted left $\mu$ times and row $\delta$ is shifted left $\nu$ times. Selecting the entries of a vector amounts to choosing an entry of the partion/dual partition array. The choice made is in the same position fpr each column of the DA. The following converts this observation to an algorithm. The entries of a vector used in a round selected from a DA are determined by $\lambda = a + b m + cm^2, 0 \le \lambda < km^2$. The vector $(\gamma,\pi,\delta)$ is taken from location $c$ of the $\gamma$ row, $a$ of the $\pi$ row and $b$ of the $\delta$ row. This algorithm is referred to as the doubly-parallel all-to-all algorithm. It has $KM^2/s$ rounds. If $K$ and $M$ are relatively prime it reduces to the all-to-all algorithm in \cite{Dra:swp} that takes $KM^2$ rounds. 

The doubly parallel algorithm can be employed even if $K$ and $M$ are relatively prime by finding $J<K$ and $L<M$ for which $J$ and $L$ have a common factor. As soon as an all-to-all involves $X>1$ items at every router, the cost is multiplied by $X^2$. If $K$ and $M$ are relatively prime, going from $D3(K,M)$ to $D3(ks,ms)$ for $K>ks$ and $M>ms$ will produce a doubly-parallel algorithm with fewer than $KM^2$ rounds if $KM^2<sJL^2$ because

$$\frac{JL^2}{s}\left(\frac{KM^2}{JL^2}\right)^2 < KM^2\  \textrm{ iff }\  \frac{KM^2}{s} \left(\frac{KM^2}{JL^2}\right)<KM^2 \ \textrm{ iff } \ \frac{KM^2}{JL^2}<s.$$
This generally works if $K-J$ and $M-L$ are small. For example, if $K=7$ and $M=16$, $J=5$, $L=15$, and $s=5$ then $KM^2/JL^2 =1.59$ so the doubly-parallel algorithm on $KM^2$ objects run on $D3(J,L)$ has $225 \times (1.59)^2 = 569$ rounds which is far less than $1792$ rounds on $D3(7,16)$.

Johnsson and Ho \cite{Joh:hyper} developed an all-to-all on a Boolean hypercube of $P$ processors with network time $t_wP/2$. For a set of size $n\ge P$ the time is $n^2/2P$. On $D3(ks,ms)$ the network time is $n^2/Ps.$

The doubly-parallel algorithm can be pipelined in several ways. The round schedules are 
$$
\begin{array}{lcccccccccccccccccc}
&&\hspace{-26pt}&(l&g&l&\hspace{-20pt})^s\\
&1&&\hspace{-20pt}&(l&g&l&\hspace{-20pt})^s\\
&&&&\hspace{-20pt}&(l&g&l&\hspace{-23pt})^s\\
&&&&&&&&\ddots\\
&&\hspace{-26pt}&(l&g&l&\hspace{-18pt})^s\\
&2&&\hspace{-18pt}&(l&g&l&\hspace{-20pt})^s\\
&&&&&&\hspace{-20pt}&(l&g&l&\hspace{-28pt})^s\\
&&&&&&&&&&\ddots
\\
\\
&&\hspace{-23pt}&(l&g&l&\hspace{-16pt})^s\\
&3 &&&&\hspace{-20pt}&(l&g&l&\hspace{-23pt})^s\\
&&&&&&&&&\ddots
\end{array}
$$
The first is a cost one schedule, the second is cost 2, and the third is cost $3$. It is obvious that Schedules $2$ and $3$ can be used without link contention. Therefore, an all-to-all algorithm runs in time $2KM^2/s$ or $3KM^2/s$ if $K=ks$ and $M=ms$.

Clearly, there is a potential for intraround conflicts between every other row in schedule $1$. Schedule $1$ can be used because of the care with which $\pi$ and $\delta$ were chosen. Suppose row $\pi$ of the original DA is left-shifted $\mu$ places. Let $t'=t+\mu \bmod s$. Column $t$ of the new DA contains $\{t'+0,t'+s,\dots , t'+(m-1)s\}$ in row $\pi$. Suppose $a$ determines which element is to be selected from each cell of row $\pi$. After the shift $t'+as$ is selected from column $t$. That is, the set of $s$ values of $\pi$ being selected for a round of the all-to-all algorithm is $\{t'+a,t'+a+1,\dots,t'+a+s-1\}$. This is a cell of the dual partition of the coset partition of $s$ in $\boldsymbol{Z} \bmod M$. This is also true of the set of $\delta s$ in a round. Therefore, a conflict of $\delta(i)$ with $\pi(i+2)$ in Schedule $1$ means that the set of $\pi's$ of round $i$ is equal to the set of $\delta s$ of round $i+2$. A single delay eliminates the conflict.

Here is an example of what happens using the earlier example for $\pi$ and $\delta$. $\gamma$ can be ignored because the only possible conflict does not involve $\gamma$. Suppose $(\mu,\nu)=(0,2)$ and $(a,b)=(1,2)$. The DA is

$$
\begin{array}{lcccl}
i&0&1&2\\
\pi&\{0,3,6,\hspace{1.5mm}9,12\}&\{1,4,7,10,13\}&\{2,5,8,11,14\}\\
\delta &\{2,5,8,11,14\}&\{0,3,6,\hspace{2mm} 9,12\}&\{1,4,7,10,13\}
\end{array}
$$
\vspace{.5cm}

\begin{singlespace}
	$$
	\begin{array}{lcccccccc}
	(a,b)&&&3 \textrm{ vectors in round } (\mu,\nu,a,b)&\vspace{5mm}\\
	(1,2)&\left(\begin{matrix} \pi\\\delta\end{matrix}\right)=&\left(\begin{matrix}{3}\\{8}\end{matrix}\right)&\left(\begin{matrix}{4}\\{6}\end{matrix}\right)&\left(\begin{matrix}{5}\\{7}\end{matrix}\right)\\
	\\
	(2,2)&&\left(\begin{matrix}{6}\\{8}\end{matrix}\right)&\left(\begin{matrix}{7}\\{6}\end{matrix}\right)&\left(\begin{matrix}{8}\\{7}\end{matrix}\right)\\
	\\
	(3,2)&&\left(\begin{matrix}{9}\\{8}\end{matrix}\right)&\left(\begin{matrix}{10}\\{6}\end{matrix}\right)&\left(\begin{matrix}{11}\\{7}\end{matrix}\right)\\
	\\
	(4,2)&&\left(\begin{matrix}{12}\\{8}\end{matrix}\right)&\left(\begin{matrix}{13}\\{6}\end{matrix}\right)&\left(\begin{matrix}{14}\\{7}\end{matrix}\right)&&b=a+2\bmod M
	\end{array}
	$$
\end{singlespace}
Note that $\pi$ in row $(2,2)$ and $\delta$ in round $(4,2)$ take the same set of values. In Schedule $1$ this causes a conflict, actually $3$ conflicts. However, each row of vectors is sent simultaneously so a one hop delay resolves the conflict. The condition $b=a+2 \bmod m$ occurs $m$ times. Therefore, for each DA a delay occurs $km$ times in $km^2$ rounds. There are $s^2$ DA's in the algorithm, so in $km^2s^2 = KM^2/s$ rounds there are $kms^2=KM$ delays. Some delays will be successive as rows $(1,2)$ and $(3,2)$ demonstrate. Schedule $1$ can only be used if $s \leq M/2$ because every round uses $2s$ local links.

The preceding discussion has proven the following:
If $K=ks$ and $M=ms$, then 

Using Schedule $1$, if $s \le M/2$ the doubly-parallel algorithm takes time $((KM^2/s + KM)/s)t_w$. 
Using Schedule $2$ the doubly-parallel algorithm takes $(2KM^2/s)t_w$ and is conflict free. 
Using Schedule $3$, the doubly-parallel algorithm takes time $3KM^2t_w/s$ and is conflict free.

A $MDF(K,M)$ has $(KM+1)M$ routers. $M$ and $(KM+1)$ are relatively prime so an algorithm like this is not possible. A partially populated Dragonfly with $KM^2$ groups may be able to exploit the idea.

If $s=1$ this theorem reduces to an algorithm originally occurring in \cite{Dra:swp}. The algorithm can be implemented on a fully populated Dragonfly provided the connectivity of groups in the Dragonfly is done properly \cite{Dra:fine}.

\section{$D3(2^k,2^m)$ and The Swapped Boolean Hypercube}

The Abelian groups used to enumerate routers and ports do not have to be cyclic groups. For example, if $M=16$ the group can be $\oplus \mathbb{Z} \bmod 2$ which is Boolean algebra on the $4$ bit quantities. If $c$ and $p$ do not use the same kind of group, the condition that $M$ and $K$ have a common factor is replaced by the condition that the group used for $c$ and the group used for $p$ have subgroups of the same size.

The Swapped Boolean Hypercube, $SBH(k,m)$, is a graph with $2^{k+2m}$ nodes. It's address space is a set of $k+2m$ long bit strings that are partitioned into three fields $(c,d,p)$. The field $c$ is the high order $k$ bits, $p$ is the low order $m$ bits and $d$ is the middle $m$ bits. If $p$ and $p'$ differ by one bit $i, (c,d,p)$ is connected to $(c,d,p')$. This link is denoted $\pi_i$ and is called a local link. If $c$ and $c'$ differ by bit $i$, there is a link denoted $\gamma_i$ connecting $(c,d,p)$ to $(c',p,d)$. Note  the swap of $p$ and $d$. There is one additional link which connects $(c,d,p)$ to $(c,p,d)$. It is denoted $Z$. If $p=d$ there is no link $Z$. All links are bidirectional. The nodes of $SBH(k,m)$ are of degree $k+m$.

The bit exchange between $(c,d,p)$ and another node depends upon the field the bit is in. The table gives the bit exchanges.
$$
\begin{array}{llc}
field&\\
c & (c,d,p)\stackrel{\gamma_i}\longrightarrow(c',p,d)\stackrel{Z}\longrightarrow(c',d,p)\\
d & (c,d,p)\stackrel{Z}\longrightarrow(c,p,d)\stackrel{\pi_i}\longrightarrow(c,p,d')\stackrel{Z}\longrightarrow(c,d',p)\\
p & (c,d,p)\stackrel{\pi_i}\longrightarrow(c,d,p')	
\end{array} 
$$
If $d=p, \gamma_i$ connects $(c,d,d)$ to $(c',d,d)$ and $Z\circ\pi_i$ connects $(c,d,d)$ to $(c,d',d)$. $SBH(k,m)$ is a dilation three emulation of the hypercube of dimension $k+m.$ It's diameter is $2k+m+3m=2k+4m$. It's nodes are degree $k+2m$. The average dilation is less than two. A node attached to $(c,d,p)$ translates a program designed for the $(k+2m)$ Boolean hypercube into a program on the Swapped Hypercube using the above paths.

If $D3(2^k,2^m)$ is constructed using Boolean arithmetic, it obviously contains $SBH(k,m).\ \gamma_i = c\oplus c', \pi_i=p\oplus p'$ and $Z$ is the $0$ global port. On $SBH(k,m)$, these paths are used only for pairs that differ by a single bit. Both directions can occur simultaneoursly. All three exchanges are vector paths so can occur simultaneously because of property 4 of Swapped Dragonflies. That is why $\pi_i$ is a $glg$ path instead of an $lgl$ path. Therefore, an ascend-descent algorithm can be performed at twice the cost of the algorithm on a $(k+2m)$ Boolean hypercube because the average dilation of the emulation is two.

If $k$ is even, $2^k$ is a square so linear algebra can be performed efficiently. If $k$ is not even, $k-1$ is and $D3(2^{k-1},2^m)$ can be found inside $D3(2^k,2^m)$. 

Johnsson and Ho \cite{Joh:hyper} developed an all-to-all algorithm on a Boolean hypercube that takes network time $t_wn/2$. The algorithm is parallel over all hypercube links. The algorithm can be run on $SBH(k,m)$ or on $D3(2^k, 2^m)$. On both a hypercube link is one, two, or three network hops. The average is two.   The algorithm takes $(2/3)(2^{k+2m}/2)t_w$ time on $SBH(k,m)$. 

The object of the following discussion is to prove that the doubly-parallel all-to-all on $D3(2^k,2^m)$ is faster than the Johnson and Ho all-to-all on $SBH(k,m)$. There is a constraint on the doubly-parallel all-to-all, $s \le M/2$, because paths on $D3(2^k,2^m)$ use two local hops. Therefore $s=min(2^k,2^{m-1})$ so the doubly-parallel algortithn takes

$$2^{k+2m}/min(2^k,2^{m-1}) = max(2^m,2^{k+m+1})t_w$$ 
time on $D3(2^k,2^m)$. This is less than $(2^{k+2m}/3)t_w$ because $m$ and $k$ are not zero. Therefore, the doubly-parallel algorithm is superior.

\section{The Broadcast Swapped Dragonfly}
The $M$ routers on a drawer of $D3(K,M)$ contain $M$ depth four edge-disjoint spanning trees
\begin{equation}
	(c,d,p)\stackrel{G} \longrightarrow (*,d,p) \stackrel{L}\longrightarrow (*,p,*) \stackrel{0} \longrightarrow (*,*,p) \stackrel{L}\longrightarrow (*,*,*).\label{eq:four}
\end{equation}

Replacing $p$ by $p'$ leads to edge disjoint paths. This fact can be used to do multiple broadcasts from source $(c,d,q)$ by starting with $(c,d,q) \rightarrow (c,d,p)$ directing each $(c,d,p)$ to do a different broadcast\footnote{This idea originated with Johnsson and Ho \cite{Ho:hyper}}. This requires five router hops for each broadcast vs three router hops for the depth three spanning tree 
$$ (c,d,p) \longrightarrow (c,d,*) \longrightarrow (*,*,d) \longrightarrow (*,*,*)$$
at $(c,d,p)$. 

Implementation of this idea requires that routers can be equipped with a program that does not depend upon their position in \ref{eq:four}.The program proposed here depends upon packets having a synchronizing header. A Swapped Dragonfly using these headers is called a Broadcast Swapped Dragonfly. The header has four entries $[b;\gamma,\pi,\delta]; b$ is a counter , $\gamma$ is a global port and $\delta$ and $\pi$ are local ports. A router interprets the header in the following way.
\begin{itemize}
\item	if $b$ is odd, use local port $\delta$ and change $b$ to $b-1$, $\delta$ to $\pi$ and $\pi$ to $0$,
\item	if $b$ is even, use global port $\gamma$ and change $b$ to $b-1$ and $\gamma$ to $0$.
\end{itemize}
A packet has arrived at an edge router when $b=0$. This program is independent of where the packet is in \ref{eq:four}. If routers cannot duplicate packets, the header becomes part of the packet. It is interpreted by a node attached to the router at each hop of the path. Here are evolutions of paths when $b=3$ and $4$;
$$\begin{array}{ccccccccccc}
 &(c,d,p) &\longrightarrow &(c,d,p+\delta) &\longrightarrow &(c+\gamma,p+\delta,d) &\longrightarrow &(c+\gamma,p+\delta,d+\pi)\\
&[3;\gamma,\pi,\delta]&& [2;\gamma,0,\pi] &&[1;0,0,\pi]&& [0;0,0,0]
\end{array}$$
$$\begin{array}{cccccccccc}
&(c,d,p) &\longrightarrow &(c+\gamma,p,d) &\longrightarrow &(c+\gamma,p,d+\delta) &\stackrel{0} \longrightarrow &(c+\gamma,p+\delta,d+\pi)&\longrightarrow & (c+\gamma,d+\delta,p+\pi)\\
	&[4;\gamma,\pi,\delta]&& [3;0,\pi,\delta] &&[2;0,0,\pi]&& [1;0,0,\pi]&&[0;0,0,0]
\end{array}$$

The synchronized header is part of the packet header. It will be necessary for the packet header to have a broadcast bit to distinguish the packet from a point-to-poins packet.

If routers can duplicate packets, $M$ broadcast take time $t_s + 5t_w$ where $t_s$ is the time required to delegate broadcasts from $(c,d,p)$ to its neighbors and $t_w$ is router latency. If routers cannot duplicate packets, the time for $M$ broadcasts is $5t_s$. Using the level three spanning tree at $(c,d,p)$, the time for $M$ broadcasts is proportional to $M$.

If a large number $X$ of broadcasts are needed the comparison is $5X/M$ to $3X$ network hops which is clearly a win for the depth-four trees.
However, chaining may change the calculation. The following is an analysis of performing $X>> M$ broadcasts using pipe-lining of the level three algorithm and the level five algorithm.The depth three tree pipe-line. 
$$
\begin{array}{ll}
(c,d,p)\stackrel{l}\longrightarrow(c,d,*)\stackrel{g}\longrightarrow(*,*,d)\stackrel{l}\longrightarrow(*,*,*)\\
\hspace{.7in}(c,d,p)\stackrel{l}\longrightarrow(c,d,*)\stackrel{g}\longrightarrow(*,*,d)\stackrel{l}\longrightarrow(*,*,*)\\
\hspace{1.4in}(c,d,p)\stackrel{l}\longrightarrow(c,d,*)\stackrel{g}\longrightarrow(*,*,d)\stackrel{l}\longrightarrow(*,*,*)\\
\hspace{2.1in}(c,d,p)\stackrel{l}\longrightarrow(c,d,*)\stackrel{g}\longrightarrow(*,*,d)
\end{array}
$$
is free of conflict if $p\neq d$ so the cost is $X$ router hops.

Pipe-lining the $M$ depth-four spanning tree is more problematic; the first local hop is the delegation step.

$$
\begin{array}{ll}
(c,d,p)\stackrel{l}\longrightarrow(c,d,q)\stackrel{g}\longrightarrow(*,q,d)\stackrel{l}\longrightarrow(*,q,*)\stackrel{0}\longrightarrow(*,*,q)\stackrel{l}\longrightarrow(*,*,*)\\
\hspace{.7in}(c,d,p)\stackrel{l}\longrightarrow(c,d,q)\stackrel{g}\longrightarrow(*,q,d)\stackrel{l}\longrightarrow(*,q,*)\stackrel{o}\longrightarrow(*,*,q)\longrightarrow(*,*,*)\\
d \neq p \hspace{1.1in} (c,d,p)\stackrel{l}\longrightarrow(c,d,q)\stackrel{g}\longrightarrow(*,q,d)\longrightarrow\\
\hspace{2.65in}\uparrow\\
\hspace{2.5in} \textrm{conflict}
\end{array}
$$
	
	So chaining in pairs gives the following:
	$$\begin{array}{l}
		(c,d,p)\stackrel{l}\longrightarrow(c,d,q)\stackrel{g}\longrightarrow(*,q,d)\stackrel{l}\longrightarrow(*,q,*)\stackrel{0}\longrightarrow(*,*,q)\stackrel{l}\longrightarrow(*,*,*)\\
		\hspace{.7in}(c,d,p) \stackrel{l}\longrightarrow (c,d,q)\stackrel{g}\longrightarrow(*,q,d)\stackrel{l}\longrightarrow(*,q,*)\stackrel{0}\longrightarrow(*,*,q)\stackrel{l}\longrightarrow(*,*,*)
		\end{array}
		$$
	delivers $2$ broadcasts every $6$ router hops for a cost of $3X/M$.
	
	In broadcast mode the level three broadcast at $(c,d,p)$ has header $[3;*,*,*]$ and the level four broadcast has header $[4;*,*,*]$. The evolution of the headers in a broadcast is

$$\begin{array}{ccccccccccc}
	&(c,d,p) &\longrightarrow &(c,d,*) &\longrightarrow &(*,*,d) &\longrightarrow &(*,*,*)\\
	&[3;*,*,*]&& [2;*,0,*] &&[1;0,0,*]&& [0;0,0,0]
\end{array}$$
$$\begin{array}{cccccccccc}
	&(c,d,p) &\longrightarrow &(*,p,d) &\longrightarrow &(*,p,*) &\stackrel{0} \longrightarrow &(*,*,p)&\longrightarrow & (*,*,*)\\
	&[4;*,*,*]&& [3;0,*,*] &&[2;0,0,*]&& [1;0,0,*]&&[0;0,0,0]
\end{array}$$
respectively. In the level four  path, the global port $0$ is used by $KM$ routers. Note that the header $[2;0,0,*]$ compels a router to send point-to-point over global port $0$ and $[1;0,0,*]$ compels a local broadcast.

The Broadcast Swapped Dragonfly $D3(2^k,2^m)$ enables the emulation of the $(k+2m)$-Boolean hypercube with uniform dilation four. The pathe $c, d,$ and $p$ are given by
$$
\begin{array}{llc}
	c & [4;\gamma,0,0]\\
	d & [4:0,0,\delta]\\
	p & [4;0,\pi,0]
\end{array} 
$$
These are all four-path but they have the advantage that all paths of a given type can be followed concurrently, and also paths of different type can be followed concurrently without link conflict. The presence of a dilation four hypercube in $D3(2^k,2^m)$ means that algorithms designed for hypercubes may be compared with the algorithms designed here and the faster algorithm used. If $K$ and $M$ are not powers of $2$, $D3(K,M)$ contains an emulation of $D3(2^k,2^m)$ with $k=\log{K}$ and $m=\log{M}$.

\section{Conclusion}

It has been shown that there are three constraints on the parameters $K$ and $M$ that lead to useful algorithms:
\begin{enumerate}
\item On $D3(K^2,M)$ an $n\times n$ matrix product can be computed in network time $4(n^2/KM)t_w$.\\
\item On $D3(ks,ms)$ an all-to-all exchange can be performed in network time $(n^2/KM^2s)t_w$.\\
\item $D3(2^k,2^m)$ contains a dilation three average two emulation of the $(k+2m)$ Boolean hypercube.\\
\item Additionally, it has been shown that equipping $D3(K,M)$ with a synchronizing counter enables $n$ broadcasts in network time $(3n/M)t_w$.
\end{enumerate}

Result 2 is the only algorithm to do an all-to-all on $P$ processors in less than $P/2$ network time. Result 3 implies that an ascend-descend algorithm can be done on $D3(K,M)$ at twice the cost of doing the algorithm on a Boolean hypercube. The first three cases may apply to $D3(J,N)$ with $J \geq K$  and $N \geq M$ because $D3(J,N)$ contains an emulaation of $D3(K,M)$.

Source-vector routing is used to define the algorithms in 1, 2, and 4. It leads to algorithms devoid of interround conflicts. Source-vectors can be defined on a Maximal Dragonfly. However, a vector $(\gamma,\pi,\delta)$ generally leads to a link conflict at the third hop when it is used by two routers. This produces interround conflicts which lead to hotspots in an application. On a Dragonfly, the algorithms studied here would be used in a deflective routing environment. Vectors would be converted to destinations.
 
Note that $D3(2^{2k},2^{2m})$ is both a $D3(K^2,M)$ and a $D3(ks,km)$. It can do a $n\times n$ matrix product in $n/\sqrt{P}$ time, an all-to-all exchange in time $n^2/\sqrt{P}\min(2^k,2^m)$ and an ascend-descend algorithm at a factor two penalty over the cost on a hypercube of the same size. Clearly, Swapped Dragonflies of the type $D3(2^k,2^m)$ or of type $D3(K,2^m)$ with $K$ only slightly larger than $2^k$ can be versatile networks. The emulation of a $(k+2m)$ Boolean hypercube in $D3(2^k,2^m)$ means that algorithms designed for hypercubes may be compared with the algorithms designed here and the faster algorithm used.

\end{document}